\documentclass[aps, prb, twocolumn,floatfix, showpacs, 10pt]{revtex4}
\usepackage{amsmath}
\usepackage{amsfonts}
\usepackage{amsbsy}
\usepackage{amssymb}
\usepackage{graphicx}
\usepackage{textcomp}
\usepackage[caption=false,singlelinecheck=false]{subfig}
\usepackage{xcolor}
\usepackage{mathrsfs}
\usepackage{bm}
\usepackage{ulem}
\usepackage{float}
\usepackage[colorlinks,linkcolor=red,citecolor=blue,filecolor=magenta,
urlcolor=cyan]{hyperref}

\allowdisplaybreaks
  
\begin{document}
\title{Stacking sensitive topological phases in a bilayer Kane-Mele-Hubbard model at quarter filling}

\author{Archana Mishra}

\author{SungBin Lee}

\affiliation{Korea Advanced Institute of Science and  Technology, Daejeon, South Korea}

\date{\today}
\begin{abstract}
Layered quasi two dimensional systems have garnered huge interest both in the advancement of technology and in understanding emergent physics such as unconventional superconductivity, topological phases. In particular, the study of topological properties in some bilayer systems like transition metal chalcogenides and iridates has been the point of attraction due to comparatively strong spin orbit coupling of transition metal ions. In this paper, we analyze the topological phases induced by the interplay of electron correlation and spin orbit coupling in different stacking orders of bilayer honeycomb lattice at quarter filling. Considering the two most common stacking orders, AA and bernel (AB) stacking, we show that the stacking order plays a crucial role in the topological phase transitions of the bilayer interacting system.
For AA stacking case, the system realizes quantum spin Hall insulator in the presence (absence) of time reversal symmetry and magnetically ordered insulator. For bernel stacking case, however, additional phases such as charge ordered normal insulator or Chern insulator with both charge and magnetic order can be stabilized. Based on our analysis, we discuss the scope of experimental realization in bilayers of transition metal chalcogenides.
%
\end{abstract}
\maketitle
\section{\label{sec:I}Introduction}
After successful synthesis of the monolayer graphene \cite{geim2007}, the family of 2D materials has broadened appreciably with a wide range of examples like hexagonal boron nitride, exfoliation of layers from transition metal chalcogenides, transition metal oxides etc. \cite{geim2013,novoselov2016}. This has lead to the realization of numerous exciting electronic and magnetic properties like quantum (spin) Hall effect,  superconductivity, spin and charge density waves etc. \cite{ando1982,haldane1988,kane2005,kane2005_1,fu2007,qi2010,xu2013,fiori2014,ajayan2016,novoselov2016,sivadas2015,chittari2016,lin2016,lee2016,park2016}. The search of these interesting phases has also been extended  to the bilayer systems \cite{ohta2006,mccann2006a,min2007,castro2007,ramasubramaniam2011,goossens2012,woods2014,liu2016,zhang2018,prada2011,qiao2011,zhang2011,xu2013,huang2018,yankowitz2018,po2018,you2018,cao2018,roy2018,wu2018,volovik2018,peltonen2018}. 
Bilayer systems share similar fascinating properties of their monolayer partner such as electrical and thermal conductivity, changing carrier density through gating or doping, mechanical stiffness etc \cite{neto2009,novoselov2012,mccann2013}. However, numerous emergent phenomena like proximity effect, charge transfer, surface reconstruction also arise due to their layered structure and this makes the system distinct from the monolayer case\cite{woods2014,novoselov2016,liu2016}. 
In particular, it turns out that a plethora of exciting topological phases emerge which very sensitively depend on the stacking patterns of the bilayers\cite{prada2011,zhang2011,qiao2011,xu2013}. 

 The bilayer compounds with heavy elements such as transition metals, can have comparatively strong spin orbit coupling (SOC) strength. SOC in these van der Waals materials with 3d-5d transition metal ions could give rise to spin Hall effect and valley Hall effect, providing a ground to study the topological phases and their phase transitions in bilayer systems \cite{shitade2009,xiao2011,xu2014,ubrig2017}. 
In addition to the effect of strong SOC, electron correlation can further give rise to unique phases such as charge and spin density wave, superconductivity, fractional Chern insulators, quantum anomalous Hall insulator etc\cite{yang2011,huang2017,liu2017,gong2017,fatemi2018}. 
Since bilayer van der Waals materials with heavy elements have significant electron correlations and SOC, the prospect of interplay of these factors in the system can lead to emergent phases with spontaneous symmetry breaking accompanied with non-trivial topology and thus, is an interesting aspect to study\cite{huang2017,barrera2018}.

%
In this paper, we study how the stacking order plays a crucial role and lead to new topological phases in bilayer honeycomb lattice (BHL) when both SOC and electron interactions are present. Focusing on quarter filling, we consider two common stacking orders of BHL: AA stacking and bernel (AB) stacking. We study the Kane-Mele-Hubbard model with  interlayer coupling in these stacking arrangement. 
%
Within mean field analysis, we show possible phases which are uniquely stabilized due to the interplay of electron interaction, SOC and stacking orders: (i) AA stacked bilayer -- time reversal symmetry (TRS) broken quantum spin Hall insulator (QSHI) and normal insulator with magnetic order (ii) Bernel (AB) stacked bilayer -- TRS broken QSHI with both magnetic order and charge order, magnetic Chern insulator and  charge ordered normal insulator. 
These stacking sensitive phases can be realized in bilayer van der Waals materials such as transition metal chalcogenides with strong SOC and comparable interaction strength.


We start by introducing the Kane-Mele-Hubbard Hamiltonian with interlayer hopping on BHL,
\begin{align}
\label{eq:1}
H\!=\!&-t \! \!\sum_{\langle ij\rangle,l,\alpha} \! \! \Big( c^\dagger_{il\alpha}c_{jl\alpha}\!+\!h.c.\Big)
+U \sum_{i,l}n_{il\uparrow}n_{il\downarrow}
\\
&
\!-\! i\lambda_{so}\! \! \sum_{\langle\langle ij\rangle\rangle,l,\alpha} \!\! \Big( c^\dagger_{il\alpha}\nu_{ij}^\alpha c_{jl\alpha}\!+\!h.c.\Big)
-t_l \!\sum_{i,l,\alpha}c^\dagger_{il\alpha}c_{i(1-l)\alpha} \nonumber
\end{align}
where $t,~\lambda_{so},~t_l$ and $U$ are the intra-layer nearest neighbor hopping, intralayer next nearest neighbor spin orbit coupling strength, interlayer hopping and onsite Coulomb repulsion respectively. $c_{il\alpha}(c^\dagger_{il\alpha})$ is the electron annihilation (creation) operator at site $i$ and layer $l$ with spin $\alpha\in{\uparrow,\downarrow}$. $l$ can have values $0$ and $1$ representing the lower and upper layers respectively. $n_{il\alpha}$ is the number density operator, $\langle ij \rangle$ and $\langle\langle ij\rangle\rangle$ represent pair of nearest neighbor sites and next nearest neighbor sites respectively. $\nu_{ij}^\alpha\!=\!-\nu_{ji}^\alpha\!=\! \pm (-1)^\alpha$, depending whether the electron is traversing from $i$ to $j$ makes a right $ +(-1)^\alpha$ or a left $-(-1)^\alpha$ turn with spin $\alpha$. We set $t=1$ in this paper.

For non-interacting system, the Hamiltonian in Eq.\eqref{eq:1}, with $ U\!=\!0$ can be written as $H_0\!=\! \sum_{\bm k}c^\dagger_{\bm{k}}h(\bm k)c_{\bm{k}}$ where the Hamiltonian matrix  $h(\bm{k})$ at a given momentum ${\bm{k}}$ is,
\begin{align}
h(\bm{k})\!=\! \sum_{i=1}^2 \Big(d_i (\bm{k})  s^0 \!\times\! \tau^0 \! \times\!\sigma^{i} \Big) \!+\! d_3(\bm{k})s^{3} \!\times\!\tau^0 \! \times\! \sigma^3 \!+\! g_l(\bm{k}). 
\end{align}
The Pauli matrices $\bm{\sigma}(\sigma^x,\sigma^y,\sigma^z)$, $\bm {\tau}(\tau^x,\tau^y,\tau^z)$ and $\bm{s}(s^x,s^y,s^z)$ are representations for ($A,B$) sublattices, ($0,1$) layers and ($\uparrow,\downarrow$) spins respectively.
$d_1(\bm{k})\!=\!1\!+\! \cos{k_1}\!+\!\cos{k_2}$, $d_2(\bm{k})\!=\!\sin{k_1}\!-\!\sin{k_2}$ and $d_3(\bm{k})\!=\!2\lambda_{so}(\sin{k_1}\!+\!\sin{k_2}\!-\!\sin{(k_1\!+\!k_2)})$. $k_1$ and $k_2$ are the momentum components along the basis vectors of honeycomb lattice $\bm e_1$ and $\bm e_2$ respectively. 
For AA stacked BHL, every sites of both layers are vertically aligned, while for the bernel stacked BHL, one of the sublattices in each layer is vertically aligned and the other sublattice has no counterpart (See Figs.~\ref{fig:2} and \ref{fig:3}). 
Thus, the interlayer hopping term is represented as $g_l(\bm{k})\!=\!t_l\, s^0\times \tau^x\times\sigma^0$ for AA stacked BHL while $g_l(\bm{k})\!=\!t_l\,s^0\times(\tau^x\times \sigma^x\!-\!\tau^y\times \sigma^y)/2$ for bernel stacked BHL. 
Half filling case (four electrons per unit cell) has been extensively studied for both AA and bernel stacked BHL\cite{prada2011}. In the absence of SOC and interactions, the bernel stacked (AA stacked) BHL contain Fermi points (circular Fermi line). For non-zero SOC, bernel stacked BHL becomes a weak topological insulator, whereas, {AA stacked BHL realizes  several phases such as metal, normal insulator and topological insulator as a function of interlayer hopping strength}\cite{prada2011}. In addition, it has been studied that electron correlation leads to different types of antiferromagnetic insulators unique to the type of stacking order. 
\cite{rozhkov2013,tao2014}. 

 Here we focus on the quarter filling of the system: two electrons are occupied per unit cell i.e., the two lowest energy bands are filled among eight bands. It turns out that the band properties at this filling also sensitively depend on the stacking order similar to the half-filling case. For non interacting case, $U=0$, increase of interlayer hopping induces the system to stabilize a topological insulator for AA stacking, whereas it becomes a trivial insulator for bernel stacked case in the presence of SOC. (A detailed review of the non-interacting BHL is given in Appendix \ref{app:1}.) 
In the presence of interactions, we analyze the Hamiltonian in Eq.~\eqref{eq:1} with $U\!\neq\!0$ within the Hartree Fock mean field approximation. 
The mean field Hamiltonian can be written as,
\begin{align}
\label{eq:3}
H_{MF}&\!=\!H_0\!+\! \sum_{i,l}\big{(}\frac{1}{2}\Delta_{il}n_{il}\!-\!\bm{M}_{il}.\bm{S}_{il}\big{)}
\!-\! \frac{1}{4U}\sum_{i,l}\big{(}\Delta^2_{il}\!-\!\bm{M}^2_{il}\big{)},
\end{align}
 where $H_0$ is the non-interacting part of the Hamiltonian in Eq.~\eqref{eq:1}. The self consistency equations are,
$\Delta_{il \alpha }\!=\!U\langle n_{il \alpha} \rangle$ and  
$\bm{M}_{il}\!=\!2U\langle\bm{S}_{il}\rangle$,
where $\Delta_{il}\!=\!\Delta_{il\uparrow}+\Delta_{il\downarrow}$ and $\bm M_{il}$ represent the total charge order parameter and magnetization respectively at site $i$ in $l$-th layer. $\bm{S}_{il}= c^\dagger_{il\alpha}\bm{\sigma}_{\alpha\beta}c_{il\beta}$ is the spin operator and $n_{il}=n_{il\uparrow}+n_{il\downarrow}$ is the total charge density at position $i,l$ with $\alpha,\beta\in{\uparrow,\downarrow}$.
Due to the absence of nesting in the Fermi surface, we neglect any kind of spin or charge density wave stabilized with finite momentum and assume the order parameters within a unit cell: scalar parameters for charge order, $\Delta_{al}$ and vector parameters for magnetic order, $\bm M_{al}$, where $a\in{A,B}$ in each layer $l\in{0,1}$.


When layers are decoupled i.e., $t_l\!=\!0$, the system is just two copies of monolayer honeycomb lattice at quarter filling which authors have studied in Ref.\onlinecite{mishra2018}. Above a critical interaction $U_c$ and a finite $\lambda_{so}$, each layer is ferromagnetically ordered along $z$ direction and thus, the ground state is doubly degenerate in bilayer system at quarter filling: magnetic orders of both layers are aligned (a) in the same direction, (b) in the opposite direction. Accordingly, the Chern insulator with total Chern number -2  is stabilized in the case (a) and the time reversal symmetry broken quantum spin Hall insulator is stabilized in the case (b). Such phases stabilized at $t_l\!=\!0$ are marked with thick black lines in both Figs.~\ref{fig:2} (a) and \ref{fig:3} (a).
\subsection{AA stacking}
\begin{figure}
\includegraphics[scale=0.9,trim=20mm 0mm 10mm 0,clip]{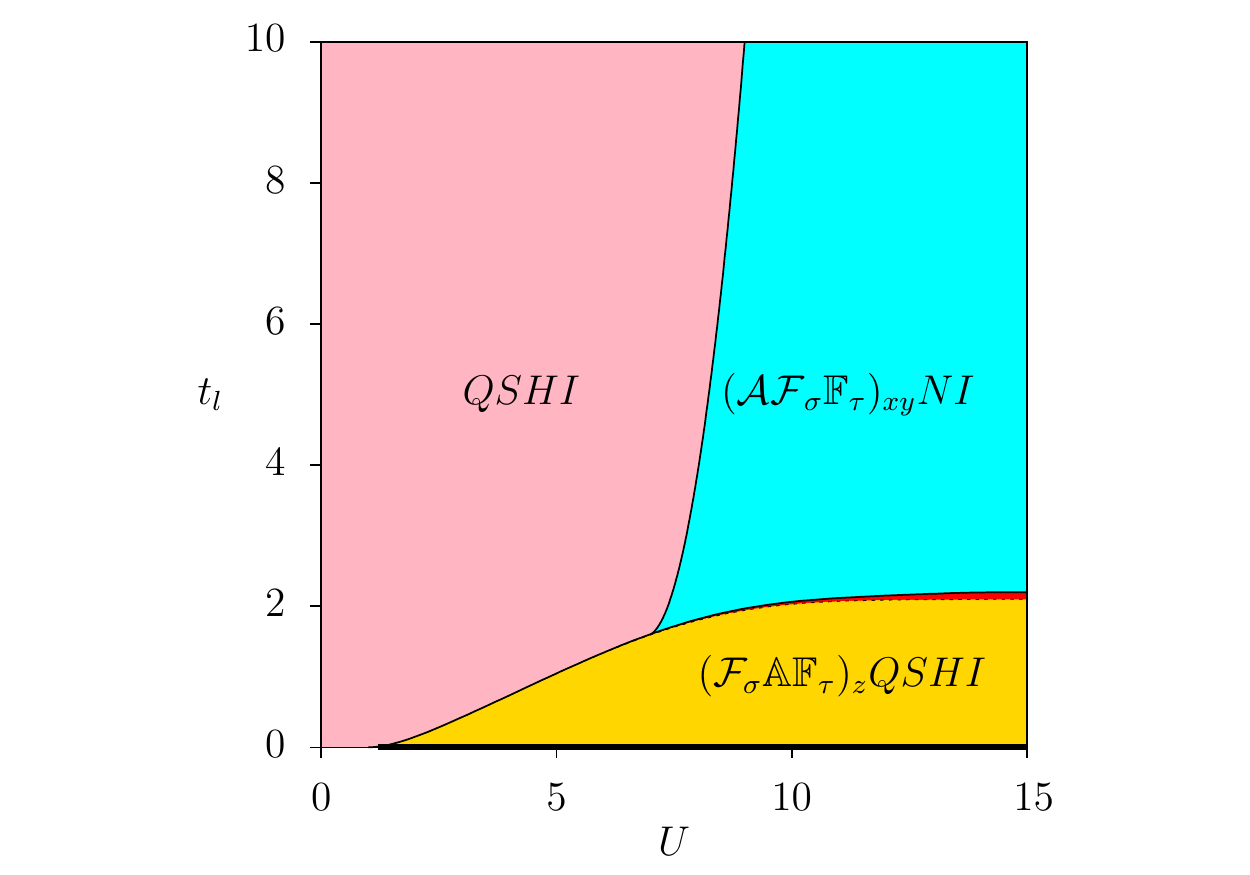}
\includegraphics[scale=0.07,trim=175mm 0mm 10mm 0mm,clip]{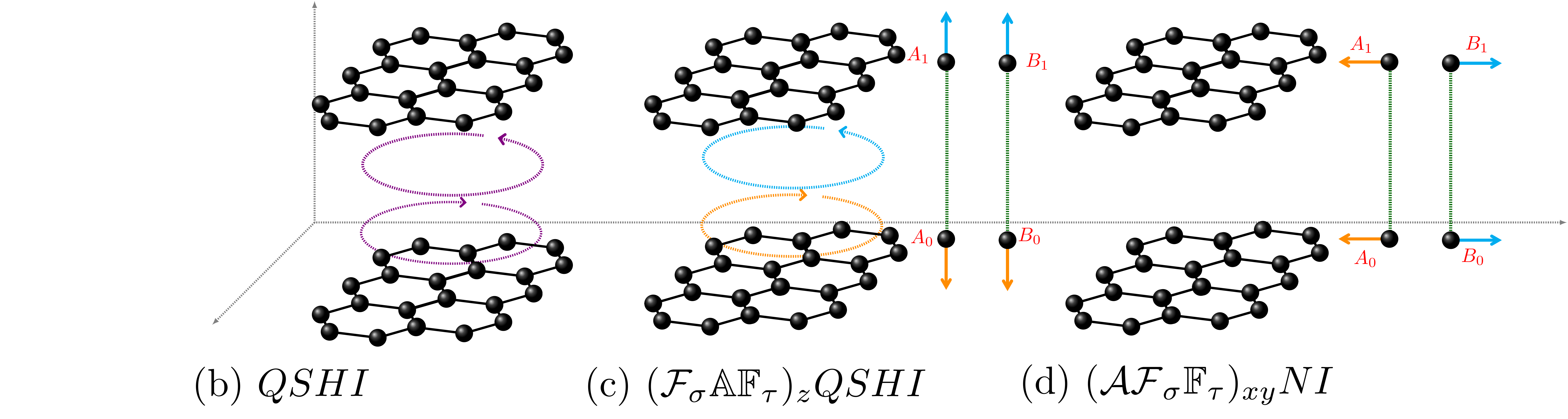}
\caption{(a) Phase diagram as functions of $U$ and $t_l$ at $\lambda_{so}=0.6$ for  AA stacking. $QSHI$ and $NI$ represents quantum spin Hall insulator and normal insulator respectively. $\mathcal{F}_\sigma (\mathbb{F}_\tau),~\mathcal{AF}_\sigma (\mathbb{AF}_\tau)$ : Ferromagnet and antiferromagnet intralayer (interlayer) ordering. The subscript ($z$, $xy$) gives the direction of the magnetization.  Solid (dashed) lines represent the second (first) order phase transitions. Schematic picture of possible phases in AA stacked BHL, (b) $QSHI$ phase,  (c) $(\mathcal{F}_{\sigma}\mathbb{AF}_{\tau})_zQSHI$ phase and (d) $(\mathcal{AF}_{\sigma}\mathbb{F}_{\tau})_{xy}NI$ phase with arrangement of magnetization and edge spin current. The straight arrows show the directions of magnetization on the sublattices. The green lines show the connectivity between the layers. Edge spin current is shown by circle arrows.}
\label{fig:2}
\end{figure}

In the absence of electron interaction, the interlayer hopping leads to a mixing of the wavefunction with equal contribution from the sublattices of both layers. Thus, the quarter filled AA stacked BHL resembles half filled monolayer honeycomb lattice and the system develops a QSHI at quarter filling for non-zero $t_l$ and finite SOC, as shown in Fig.~\ref{fig:2}b (See Appendix \ref{app:1}). The phase diagram as functions of $t_l$ and $U$ is given in Fig.~\ref{fig:2}a for $\lambda_{so}=0.6$.

Increasing $U$ above a critical value, there is a transition to the phase where intralayer-ferro and interlayer-antiferromagnetic order is favored for small $t_l$. In the presence of SOC, the system prefers the magnetization along the z-direction due to SU(2) spin symmetry breaking and goes into a time reversal symmetry broken quantum spin Hall phase. We denote it as $(\mathcal{F}_{\sigma}\mathbb{AF}_\tau)_zQSHI$ where, as discussed before, $\sigma$ represents the sublattices $(A,B)$ within a layer and $\tau$ represents the $(0,1)$ layers. The schematic diagram of the $(\mathcal{F}_{\sigma}\mathbb{AF}_{\tau})_zQSHI$ phase is shown in Fig.~\ref{fig:2}c. The direction of magnetization and edge spin current arrangement is also shown in Fig.~\ref{fig:2}c. The intralayer ferromagnetism has already been studied in Ref.~\onlinecite{mishra2018} which is realized due to the  kinetic energy gained by electrons hopping with spin polarization. For small value of $t_l$, the kinetic energy gained due to the  interlayer hopping is very less and the exchange interaction favoring antiferromagnetism results in the interlayer antiferromagnetic order. This can further be understood by comparing the total mean field energies of the system in $(\mathcal{F}_{\sigma}\mathbb{AF}_\tau)$ phase (say case I) and $(\mathcal{F}_{\sigma}\mathbb{F}_\tau)_z$ phase (say case II).
For small $t_l$, the magnitudes of charge order parameter and magnetization are same on each sublattice i.e. the electrons are completely in one of the spin states (either up or down) in each of the sublattices of the unit cell.  For case I, $\Delta_{a0}=M^z_{a0}=\Delta_{a1}=-M^z_{a1}=\Delta$ and the total mean field energy for a given $\bm k$ is $E_{MF,I}(\bm k)=\Delta-\sqrt{4t^2_l+\Delta^2}-2\sqrt{d^2_1(\bm k)+d^2_2(\bm k)+d^2_3(\bm k)}$. For case II, $\Delta_{al}=M^z_{al}=\Delta$ and  $E_{MF,II}(\bm k)=-2\sqrt{d^2_1(\bm k)+d^2_2(\bm k)+d^3_2(\bm k)}$, thus, $E_{MF,I}(\bm k)<E_{MF,II}(\bm k)$. 
 On increasing $t_l$, the competition between kinetic energy via interlayer hopping and  exchange interaction increases resulting in $\Delta_{al}>|M^z_{al}|$ in the $\mathcal{F}_{\sigma}\mathbb{AF}_{\tau}$ phase. 
 
For large $U$, on further increasing $t_l$, there is another phase transition to the normal insulator with intralayer antiferro and interlayer ferromagnetic ordering and is denoted as $(\mathcal{AF}_{\sigma}\mathbb{F}_{\tau}){NI}$ phase. In this case, the kinetic energy can be gained by the interlayer hopping of electrons and thus, interlayer ferromagnetic order is favored whereas intralayer has antiferromagnetic ordering. The magnetization in this phase is along the xy-direction in the presence of SOC and it is denoted as $(\mathcal{AF}_{\sigma}\mathbb{F}_{\tau})_{xy}{NI}$ as shown in Fig.~\ref{fig:2}d. 
 {The choice of this phase can be further understood by considering the single particle energy of the mean field Hamiltonian which is explicitly discussed in Appendix \ref{app:2}. 
In between $(\mathcal{F}_{\sigma}\mathbb{AF}_{\tau})_z$ and $(\mathcal{AF}_{\sigma}\mathbb{F}_{\tau})_{xy}$ phases, there is a small region where both of these phases coexist as marked in red in Fig.~\ref{fig:2}a. It is interesting to note that in this region a non-collinear magnetic ordering is stabilized as a result of competing interlayer hopping and Coulomb repulsion.  


\subsection{Bernel (AB) stacking}
\begin{figure}
\includegraphics[scale=0.9,trim=20mm 0mm 3mm 0mm,clip]{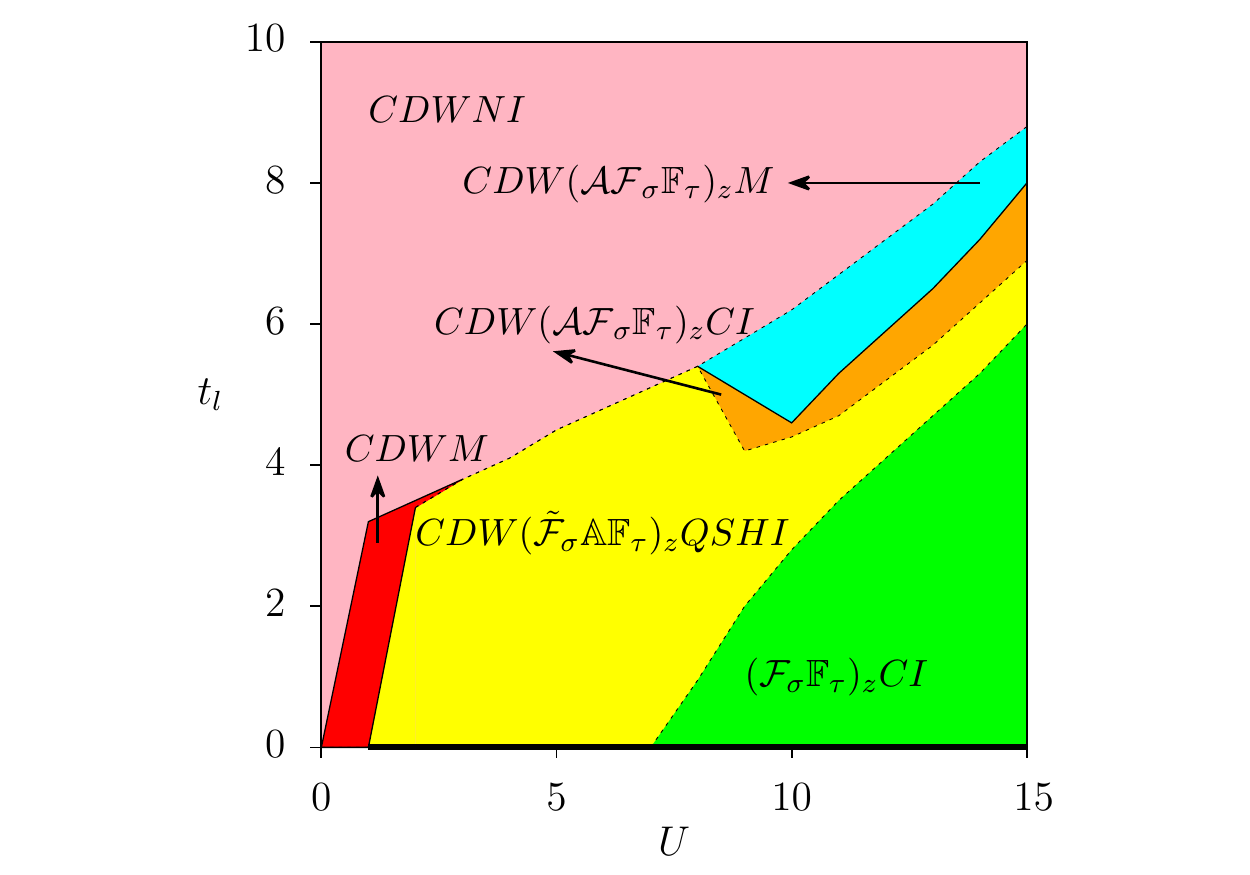}
\includegraphics[scale=0.09,trim=100mm 0mm 10mm 0,clip]{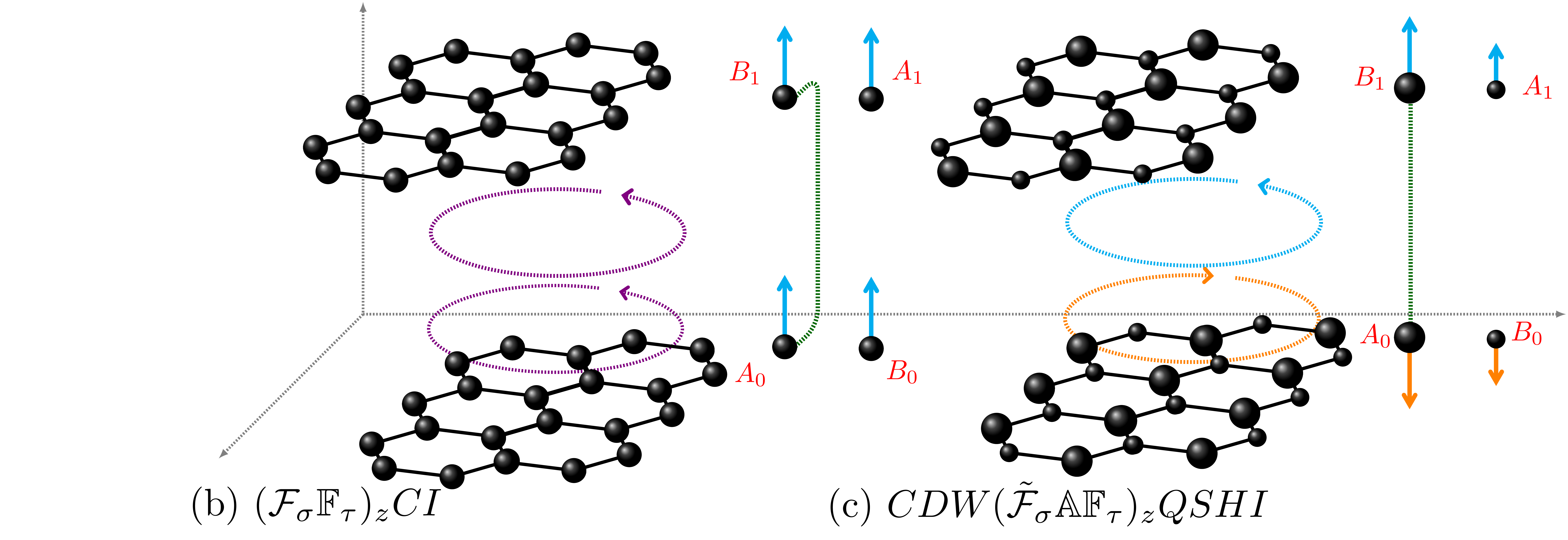}
\includegraphics[scale=0.09,trim=100mm 0mm 10mm 0,clip]{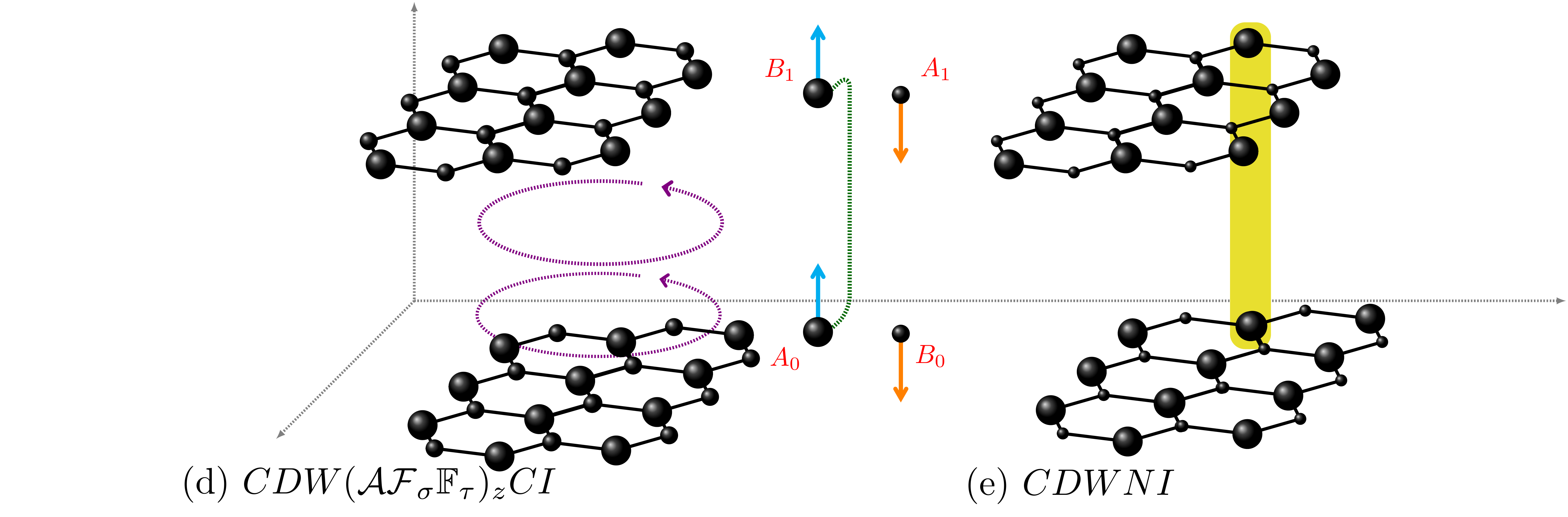}
	\caption{(a) Phase diagram as functions of $U$ and $t_l$ at $\lambda_{so}=0.6$ for  bernel stacking. $CI,~QSHI$ and $NI$ represent Chern insulator, quantum spin Hall insulator and normal insulator respectively. $\mathcal{F}_\sigma (\mathbb{F}_\tau),~\mathcal{AF}_\sigma (\mathbb{AF}_\tau)$ : Ferromagnet and antiferromagnet intralayer (interlayer) ordering. The subscript ($z$, $xy$) gives the direction of the magnetization. $CDW$ denotes charge density wave. The subscript gives the direction of the magnetization.  Solid (dashed) lines represent the second (first) order phase transitions in the phase diagram. Schematic picture of possible phases in bernel stacked BHL, (b)   $(\mathcal{F}_{\sigma}\mathbb{F}_{\tau})_zCI$, (c) $CDW(\mathcal{\tilde F}_{\sigma}\mathbb{AF}_{\tau})_zQSHI$ phases, (d) $CDW\mathcal{AF}_{\sigma}(\mathbb{F}_{\tau})_zCI$ and (e) $CDWNI$ phases with arrangement of magnetization and edge spin current. The size of the ball represents the charge density at the sites. 
		 The straight arrows show the directions of magnetization on the sublattices. The green lines show the connectivity between the layers. Edge spin current is shown by circling arrows. The yellow shade connecting $A_0$ and $B_1$ sublattices in (e) represents dimers at large $t_l$.}
	\label{fig:3}
\end{figure}

Fig.~\ref{fig:3}a gives the phase diagram for the bernel stacked BHL as a function of $t_l$ and $U$ for $\lambda_{so}=0.6$. 
For bernel stacking case, unlike AA stacking, the charge disproportionation between sublattices are naturally induced due to inequivalent interlayer hopping. For very small $U$, on increasing $t_l$ from zero, the system develops a charge density wave phase with $(\Delta_{A0}=\Delta_{B1})>(\Delta_{B0}=\Delta_{A1})$ due to the increase of kinetic energy through interlayer hopping along $A0-B1$ bond and the system stabilizes the charge density wave metallic phase ($CDWM$). The charge disproportionation between the sublattices increases with increasing $t_l$ and the system becomes a charge density wave normal insulator ($CDWNI$) at large $t_l$ limit.  (see Fig.~\ref{fig:3}e). 

%

When interaction strength is large enough to develop both charge and magnetic order, the choice of the ground state depends on the interaction strength. 
For very large $U$, the system chooses $\mathcal{F}_{\sigma}\mathbb{F}_{\tau}$  phase as the ground state as soon as the interlayer hopping is turned on. As discussed in the previous subsection, the intralayer magnetic ordering favors ferromagnetism. The interlayer ferromagnetism can be understood by comparing the total mean field energies of the two cases : $\mathcal{F}_{\sigma}\mathbb{AF}_{\tau}$ (case I) and $\mathcal{F}_{\sigma}\mathbb{F}_{\tau}$ (case II). 
Similar to the case of AA stacking, it is reasonable to consider that the magnitudes of charge order and magnetic order are equivalent for small $t_l$ and large $U$; for case I, $\Delta_{a0}=M^z_{a0}=\Delta_{a1}=-M^z_{a1}=\Delta$ and for case II, $\Delta_{al}=M^z_{al}=\Delta$. For $\lambda_{so}=0$, the total single particle energy at quarter filling at a given $\bm k$ for case I is $E_{MF,I}=\Delta-
\Big{(}2t_l^2+4(d^2_1(\bm k)+d^2_2(\bm k))+\Delta^2+2\sqrt{t_l^4+4(t_l^2+\Delta^2)(d^2_1(\bm k)+d^2_2(\bm k))}\Big{)}^{1/2}$ and for case II is  $E_{MF,II}=-\sqrt{t_l^2+4(d^2_1(\bm k)+d^2_2(\bm k))}$. For large $U$ and small $t_l$, expansion of $E_{MF,I}$ upto second order of $t_l$ is $E_{MF,I}\approx -2\sqrt{d^2_1(\bm k)+d^2_2(\bm k)}-\frac{t_l^2}{(2(d^2_1(\bm k)+d^2_2(\bm k))_+\Delta)}$.
Hence, in this regime, $E_{MF,I}>E_{MF,II}$ and $\mathcal{F}_{\sigma}\mathbb{F}_{\tau}$ phase is the ground state. In the presence of SOC, this statement still holds but the magnetization is chosen along the z-direction and the system goes into a magnetic Chern  insulating phase with $C_f=-2$. The schematic picture of the system in $(\mathcal{F}_{\sigma}\mathbb{F}_{\tau})_zCI$ phase is shown in Fig.~\ref{fig:3}b. The direction of magnetization and edge  spin current arrangement is also shown in Fig.~\ref{fig:3}b.

For intermediate $U$, charge order starts developing along with the magnetic order as $t_l$ is increased with 
 $(\Delta_{A0}=\Delta_{B1})>(\Delta_{A1}=\Delta_{B0})$. 
 The system develops an antiferromagnetic ordering between the layers and ferromagnetic ordering within the layers. The intralayer ferromagnetic ordering breaks the inversion symmetry of the system (represented as $\mathcal{\tilde{F}}$). Thus, the system is in a $CDW (\mathcal{\tilde{F}}_{\sigma}\mathbb{AF}_{\tau})_zQSHI$ phase as  shown in Fig.~\ref{fig:3}c. 
At large $U$, there is a first order transition from $(\mathcal{F}_{\sigma}\mathbb{F}_{\tau})_zCI$ to $CDW (\mathcal{\tilde{F}}_{\sigma}\mathbb{AF}_{\tau})_zQSHI$ phase as  $t_l$ is increased. 	{Similar to AA stacking, on increasing $t_l$, the competition between kinetic energy due to interlayer hopping and  exchange interaction increases resulting in $\Delta_{al}>|M^z_{al}|$ for the $\mathcal{F}_{\sigma}\mathbb{AF}_{\tau}$ phase. The constant term $-\frac{1}{4U}\sum_{i,l}(\Delta^2_{il}-\bm {M}^2_{il})$ in Eq.~\eqref{eq:3} is non-zero lowering the total energy of $\mathcal{F}_{\sigma}\mathbb{AF}_{\tau}$ phase. Thus, as  $t_l$ is increased, $\mathcal{F}_{\sigma}\mathbb{AF}_{\tau}$ phase gradually becomes lower in energy than the $\mathcal{F}_{\sigma}\mathbb{F}_{\tau}$ phase.} 

For both large $t_l$ and $U$, there is an intermediate phase with antiferromagnetic ordering within the layers ($\mathcal{AF}_{\sigma}$) and ferromagnetic ordering between the layers ($\mathbb{F}_{\tau}$) {along with the charge ordering  $(\Delta_{A0}=\Delta_{B1})>(\Delta_{A1}=\Delta_{B0})$}. {In this regime, the system tries not only to maximize the kinetic energy through interlayer hopping because of large $t_l$ but also to maximize the exchange energy for large $U$, resulting in $\mathcal{AF}_{\sigma}\mathbb{F}_{\tau}$.}
The system is a Chern insulator ($CI$) in this phase with $C_f=-2$, labeled as $CDW(\mathcal{AF}_{\sigma}\mathbb{F}_{\tau})_zCI$. The schematic picture of this phase is shown in Fig.~\ref{fig:3}d.
When $t_l$ is further increased, the system goes into a metallic regime with the same charge and magnetic ordering at large $U$. 
Hence, as shown in Fig.~\ref{fig:3}(a), one can expect several topological phase transitions when both electron interaction and interlayer hopping play a role. 

In summary, we have studied the Kane Mele Hubbard model in  bilayer honeycomb lattice with AA and bernel stacked order. Motivated by the rising research on bilayers of van der Waals materials and their magnetic and topological properties \cite{yang2011,liu2017,gong2017,fatemi2018, barrera2018,shitade2009,xiao2011,xu2014,ubrig2017,huang2017,mishra2018}, we have investigated the topological phase transitions accompanying the Landau phase transitions induced by the interplay of SOC and electron correlation.
It turns out that the realization of exotic topological phases with magnetic order and charge order are very versatile depending on the stacking order. When both SOC and electron interaction come into play, the AA stacked BHL prefers either time reversal symmetry preserved (broken) quantum spin Hall insulating phase or trivial magnetic insulator. On the other hand, for bernel stacked BHL, we find non-trivial topological phases like Chern insulator or quantum spin hall insulator where charge order and magnetic order are both present. Strong interlayer hopping in this stacking also stabilizes charge ordered insulators. 
 
Bilayer systems are a connecting link to understand  the electronic, magnetic and topological properties of 2d lattice systems from their 3d counterpart. Especially, understanding 2d magnetism from 3d bulk in van der Waals materials like transition metal trichalcogenides have been a prime area of research in both theory and experiments \cite{wiedenmann1981,joy1992,carteaux1995,wildes1998,casto2015,sivadas2015,chittari2016}. Furthermore, the topological phases and phase transitions in the layered 2d materials yet another interesting aspect of research. Our study gives a guidance to understand the interaction driven topological phase transitions and their sensitivity to the stacking, which in return can be useful to analyze the deviation in these properties as we increase the layers from 2d lattice to 3d bulk.
 In particular, the bilayers of van der Waals materials with 3d-5d transition metal ions have strong spin orbit coupling and comparable electron correlations and thus are the potential candidate for experimental realization of our study. In transition metal trichalcogenide series, for instance, the combination of SOC and crystalline electric fields can split $t_{2g}$ orbitals of transition metals ions into lower quartet orbitals with the effective total angular momentum $j = 3/2$ and upper doublet with $j = 1/2$ in the atomic limit \cite{sugano2012,enda2013}. With 18 or 22 number of electrons in $d$ orbitals per unit cell (two sites in each layer), the $j = 3/2$ orbitals are fully filled, while the $j = 1/2$ orbitals of  the unit cell have two or six electrons in total, resulting in effective quarter or three-quarter fillings with pseudospin-1/2 model. Such kind of fillings can be realized 
by mixed valence of transition metal ions in the sublattices of the unit cell possible by doping via gating or hydrogen substitution \cite{luo2019}. Hence, bilayers of various two dimensional materials under above condition would be an ideal place to realize these interesting non-trivial phases with magnetism.



\appendix
\section{\label{app:1}Non-interacting case}
In this section, we give a brief review of the non-interacting bilayer honeycomb lattice with SOC for AA and bernel stacked cases.
In the absence of interlayer hopping, the energy spectrum of BHL is same as that of the monolayer case with each band being quartic degenerate. The system is a semi-metal at half-filling with the energy bands touching at the Dirac points in the absence of SOC and a quantum spin Hall insulator in the presence of SOC. At quarter or three quarter filling, the system is a metal both in the absence and presence of SOC. 

The quartic degeneracy breaks to double degeneracy in the presence of interlayer hopping. For AA stacking, the energy eigenvalues are given as $E_{AA}=\pm t_l\pm\sqrt{d^2_1(\bm{k})+d^2_2(\bm{k})+d^2_3(\bm{k})}$.  At half filling, the bands touch when $t_l=\sqrt{d^2_1(\bm{k})+d^2_2(\bm{k})+d^2_3(\bm{k})}$. Thus, the band touching moves from the Dirac points to the $\Gamma$ point in the Brillouin zone as the interlayer hopping is increased and finally opens a gap becoming a trivial insulator. Hence, there is a transition from QSHI to trivial insulator on varying the interlayer hopping at half filling for non-zero SOC. At both quarter and three quarter filling, the bands touch at the Dirac points for $\lambda_{so}=0$ in the presence of interlayer hopping and open a gap becoming a QSHI for non-zero SOC. 

The story is different for the bernel stacked BHL. The non-interacting single particle energy eigenvalues for this stacking is given as $E_{AB}=\pm\frac{1}{2}\sqrt{\Big{[}t_l\pm\sqrt{t_l^2+4\Big{(}d^2_1(\bm{k})+d^2_2(\bm{k})\Big{)}}\Big{]}^2+4d^2_3(\bm{k})}$.  Here again, each of the band is doubly degenerate due to Kramer degeneracy for non-zero $t_l$. At half filling, the bands touch quadratically at the same momentum points ($K, K'$) which were the Dirac points for monolayer case in the absence of SOC. The system becomes a quantum spin Hall insulator in the presence of SOC. However unlike monolayer case, the Chern number for the bands near half filling below the Fermi level is -2 for spin up and +2 for spin down band. In this case, the contribution to the Chern number value is 1 per spin per valley ($K,~K'$) 
 and hence the total Chern number value is $-(+)2$ for spin up (down) case. At quarter and three quarter filling, both in the absence and presence of $\lambda_{so}$, the bands start separating as soon as the interlayer hopping is turned on and the system gradually becomes a trivial insulator. For very large $t_l$, in the absence of SOC, the lowest band per spin has zero Chern number. At quarter filling, the energy gap to the excited state is large for large $t_l$ and adding small SOC term will not close this gap. Hence, the system remains to be a trivial insulator for non-zero $t_l$ as the band crossing between the filled and empty bands occur at only $t_l=0$.

From the above discussion, we see that the topology of the bands at both half and quarter filling is dependent on the order of stacking.

\section{\label{app:2}Phase diagram explanation of AA stacked BHL for large $t_l$ and $U$}
This section gives a detailed description of the ground state for large $U$ and $t_l$ in AA stacked BHL.
In the absence of SOC, for large $t_l$, the single particle energy for $(\mathcal{F}_{\sigma}\mathbb{AF}_{\tau})$ phase is $\Delta-\sqrt{4t_l^2+m^2}-2\sqrt{d_1^2(\bm k)+d_2^2(\bm k)}$ where $m=M^{z}_{A1}-M^{z}_{A2}$. In $(\mathcal{AF}_{\sigma}\mathbb{F}_{\tau})$ phase, the single particle energy is $\Delta-2t_l-\sqrt{m^2+4(d_1^2(\bm k)+d_2^2(\bm k))}$ where $m=M^{z}_{A1}-M^{z}_{B1}$.
For large $U$ and $t_l$, we can see that the $(\mathcal{AF}_{\sigma}\mathbb{F}_{\tau})$ phase is lower in energy and hence, is the ground state of the system.
While for $\lambda_{so}=0$, the magnetization can be aligned along any direction, the magnetization is aligned along the $xy-$ direction in the presence of SOC. To understand this choice of alignment, we consider the SOC to be small but finite and calculate the perturbation to energy upto second order taking SOC term as the perturbation term. The second order correction in energy is $-\lambda_{so}^2\Big{(}d_1^2(\bm k)+d^2_2(\bm k)\Big{)}/\Big{(}m^2+d_1^2(\bm k)+d^2_2(\bm k)\Big{)}^{3/2}$ for $(\mathcal{AF}_{\sigma}\mathbb{F}_{\tau})_z$ phase with magnetization aligned along $z-$ direction. The secone order correction in energy is $-\lambda_{so}^2\Big{(}2/(m^2+d_1^2(\bm k)+d^2_2(\bm k))^{1/2}+2(d_1^2(\bm k)+d^2_2(\bm k))/(m^2+d_1^2(\bm k)+d^2_2(\bm k))^{3/2}\Big{)}$ for $(\mathcal{AF}_{\sigma}\mathbb{F}_{\tau})_{xy}$ phase with magnetization aligned along $xy-$ direction. As we see that the leading order energy correction goes as $1/m$ for $(\mathcal{AF}_{\sigma}\mathbb{F}_{\tau})_{xy}$ and $1/m^3$ for $(\mathcal{AF}_{\sigma}\mathbb{F}_{\tau})_z$ phase, hence, the phase with magnetization aligned along $xy-$ direction has lower energy than the phase with magnetization aligned along the $z-$ direction for large $U$ and $t_l$. The system has zero Hall conductivity and gapped edge states in the $(\mathcal{AF}_{\sigma}\mathbb{F}_{\tau})_{xy}NI$ phase and is a trivial insulator. 

\section{Edge state plots}
We show the band dispersion with edge states in Fig.~\ref{fig:4} and \ref{fig:6} for the phases described in the phase diagram in the main text.
The edge states for all the phases in the phase diagrams, Fig.~\ref{fig:2} and ~\ref{fig:3}, for both the stackings are investigated by considering a zigzag boundary along the y-direction but periodic boundary conditions along the x-direction. 


Fig.~\ref{fig:4} shows the edge states alongwith the bulk energy spectrum for $(\mathcal{F}_{\sigma}\mathbb{AF}_{\tau})_zQSHI$ phase  at $t_l=2$ and $(\mathcal{AF}_{\sigma}\mathbb{F}_{\tau})_{xy}NI$ phase at $t_l=5$  for $U=13$ and $\lambda_{so}=0.6$ in AA stacked BHL. 
\begin{figure}[H]
	\subfloat[]{\includegraphics[scale=0.13,trim=50mm 0mm 40mm 0,clip]{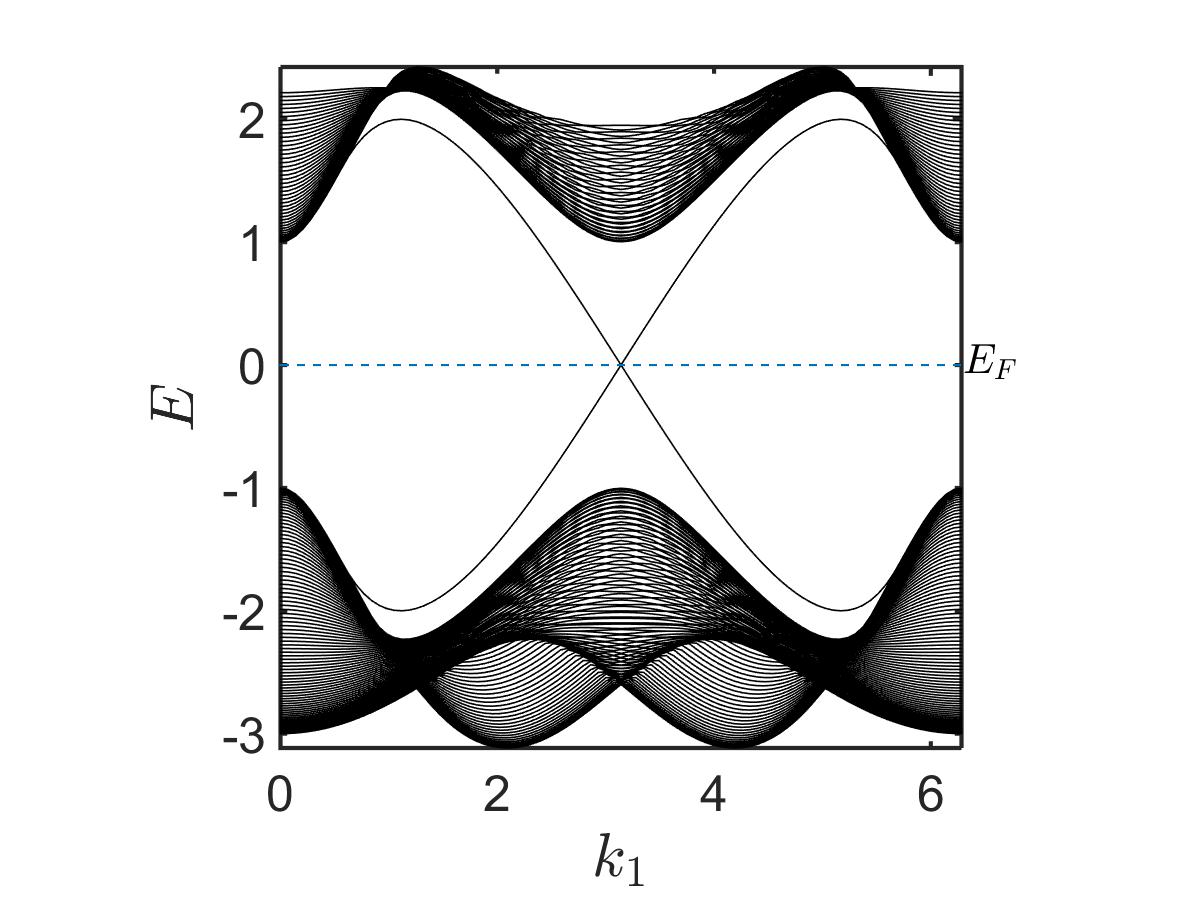}}
	\subfloat[]{\includegraphics[scale=0.13,trim=50mm 0mm 40mm 0,clip]{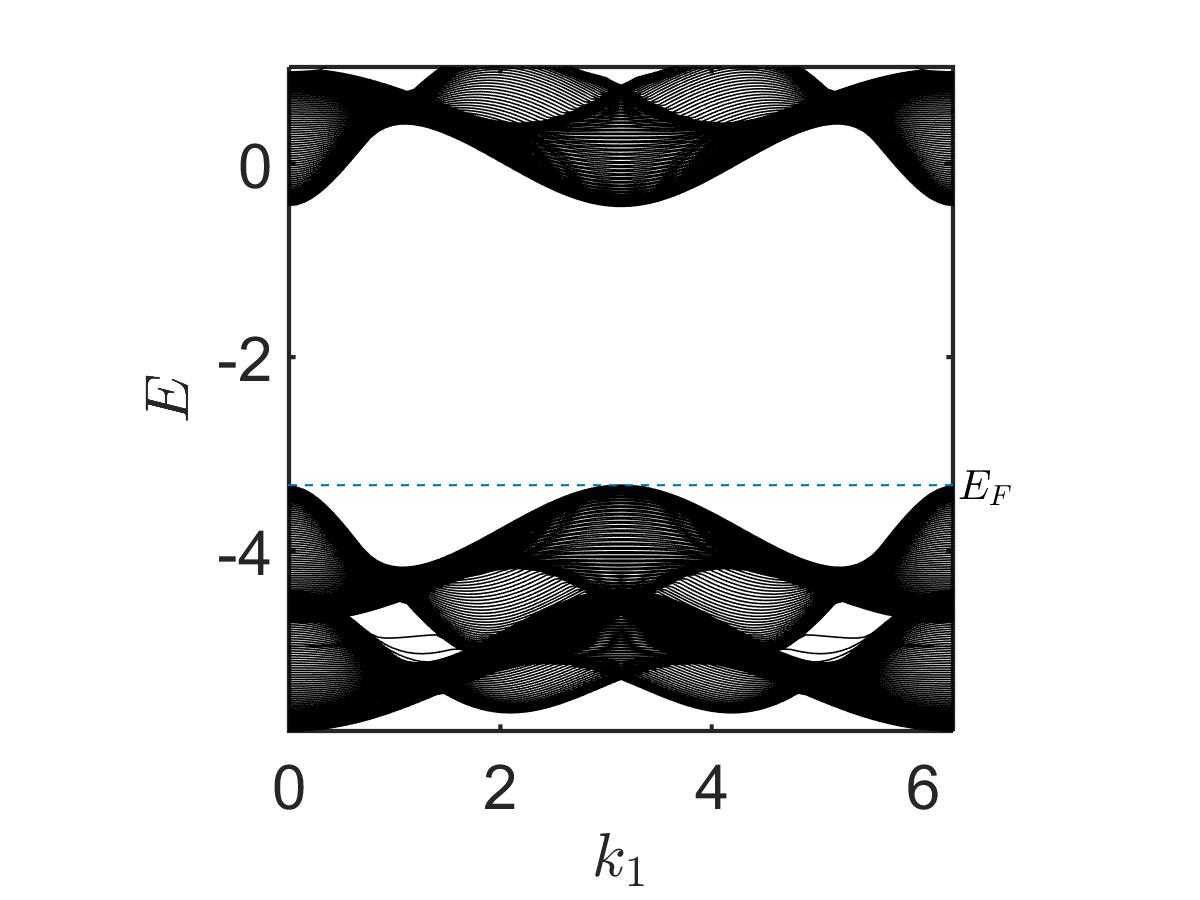}}
	\caption{\label{fig:4}}{Band dispersion of the AA stacked BHL with zigzag edge along y- direction at quarter filling for $U=13$, $\lambda_{so}=0.6$ and (a) $t_l=2$ in$(\mathcal{F}_{\sigma}\mathbb{AF}_{\tau})_zQSHI$ phase,  (b) $t_l=5$ in $(\mathcal{AF}_{\sigma}\mathbb{F}_{\tau})_{xy}NI$ phase. $E_F$ is the Fermi energy level.}
\end{figure}


\begin{figure*}
	\subfloat[]{\includegraphics[scale=0.25,trim=50mm 0mm 40mm 0,clip]{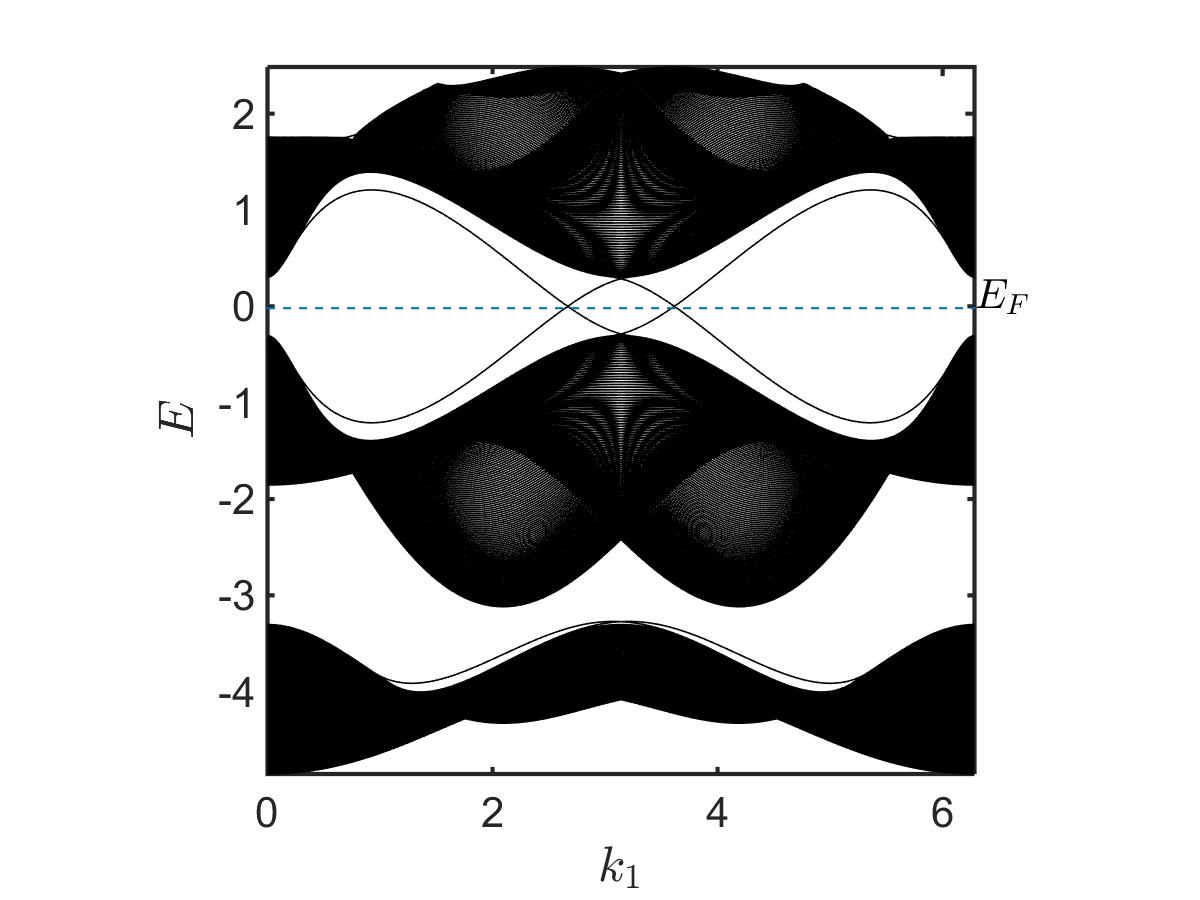}}
	\subfloat[]{\includegraphics[scale=0.25,trim=50mm 0mm 40mm 0,clip]{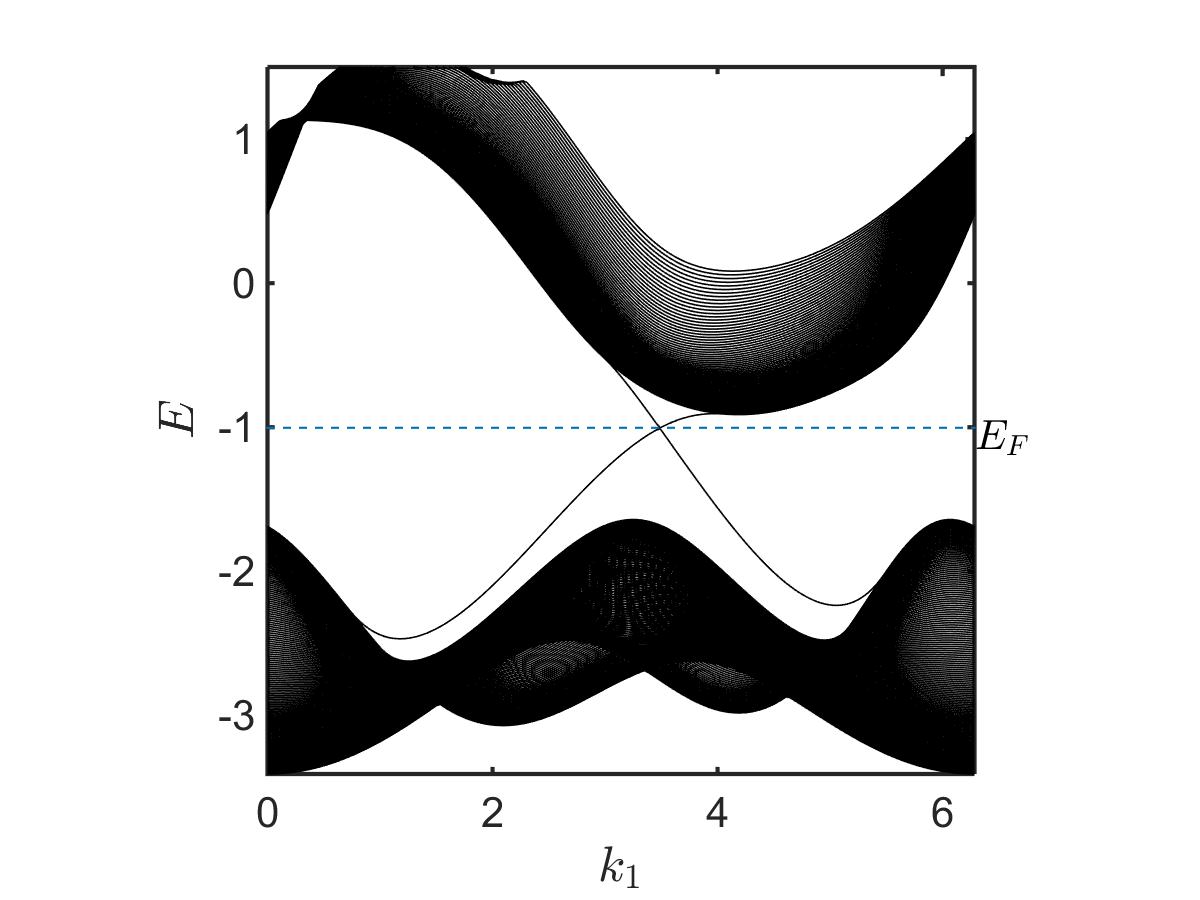}}\\
	\subfloat[]{\includegraphics[scale=0.25,trim=50mm 0mm 40mm 0,clip]{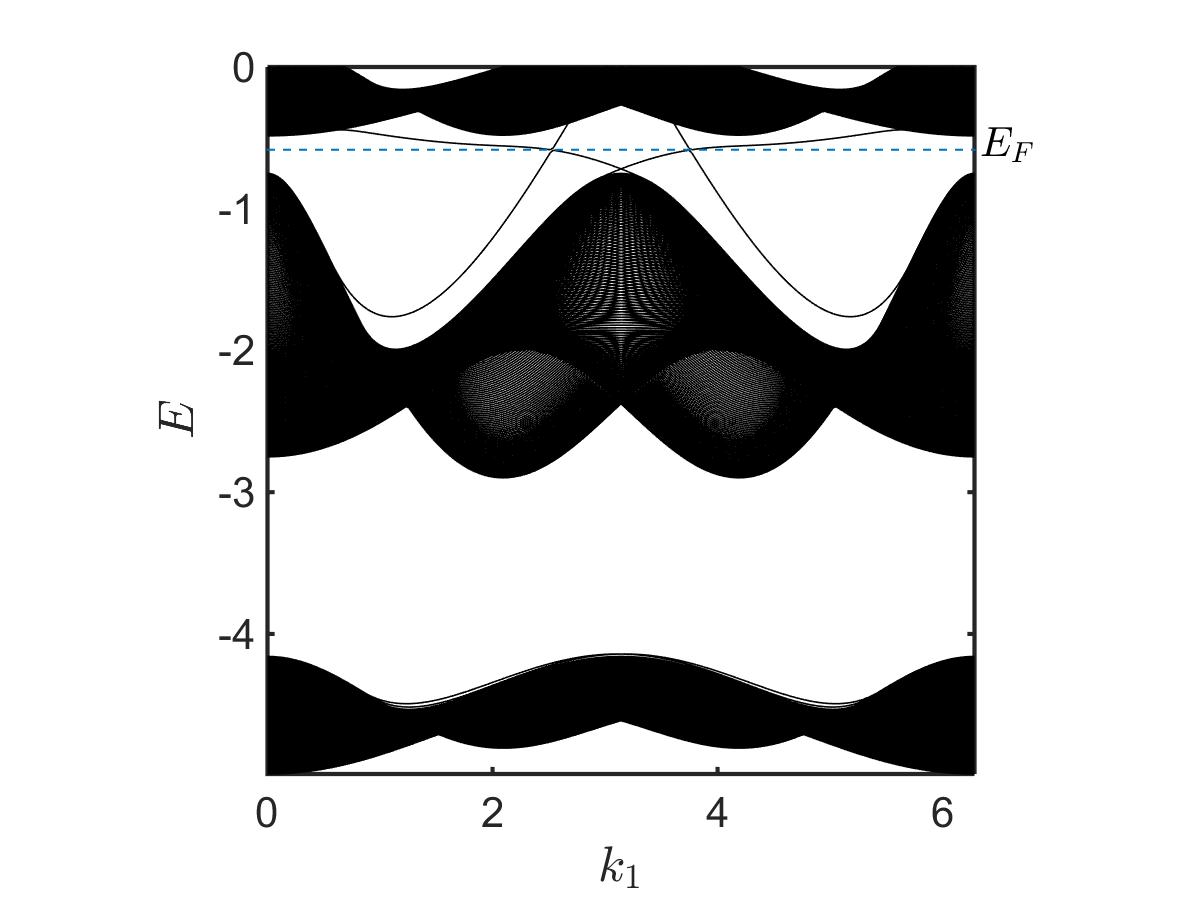}}
	\subfloat[]{\includegraphics[scale=0.25,trim=50mm 0mm 40mm 0,clip]{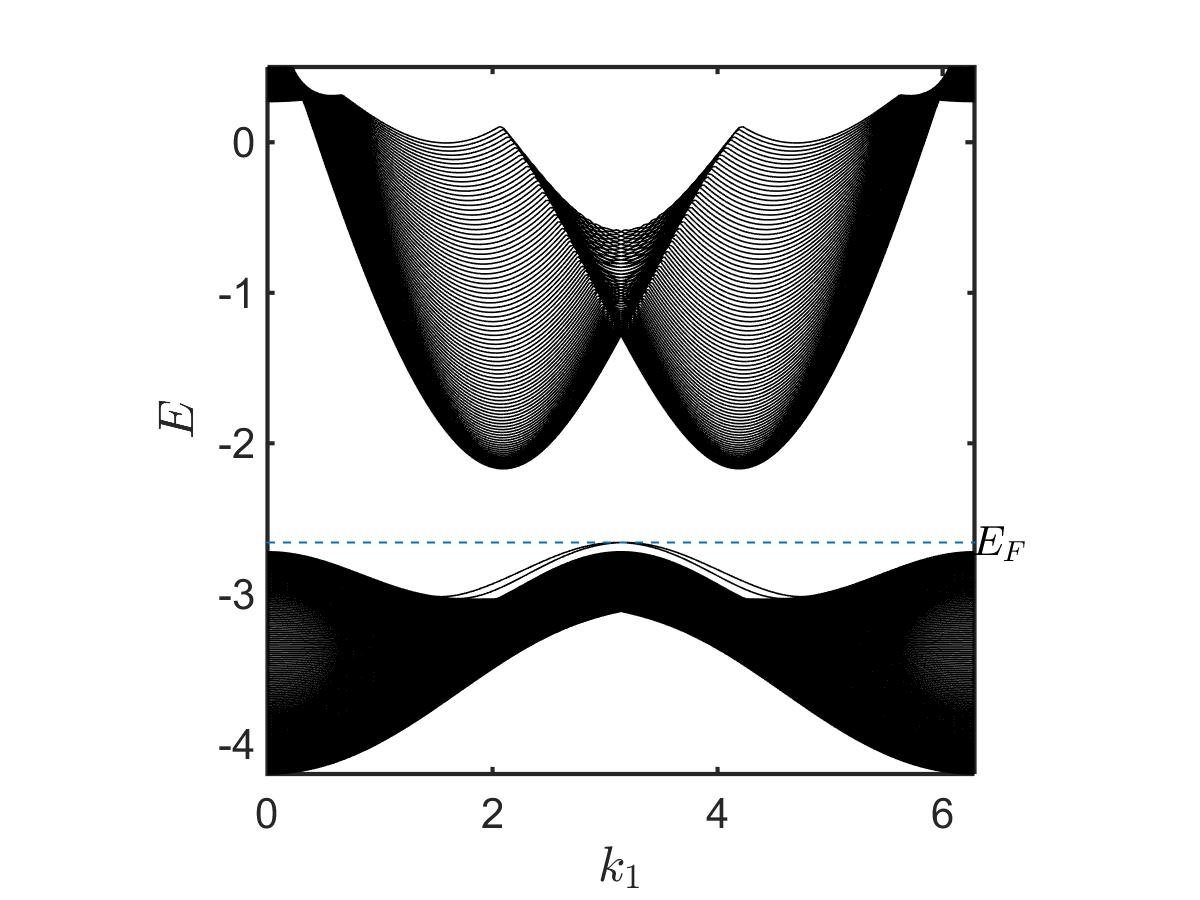}}
	\caption{\label{fig:6}}{Band dispersion of the bernel stacked BHL with zigzag edge along y- direction at quarter filling for $U=13$, $\lambda_{so}=0.6$ and (a) $t_l=3$ in $(\mathcal{F}_{\sigma}\mathbb{F}_{\tau})_zCI$  phase, (b) $t_l=5$ in $CDW~(\mathcal{\tilde F}_{\sigma}\mathbb{AF}_{\tau})_zQSHI$, (c) $t_l=6$ in $CDW~(\mathcal{AF}_{\sigma}\mathbb{F}_{\tau})CI$ and $t_l=8$ in $CDWNI$ phase. $E_F$ is the Fermi energy level.}
	\end{figure*}

 Fig.~\ref{fig:6} shows edge states for  $(\mathcal{F}_{\sigma}\mathbb{F}_{\tau})_zCI$, $CDW~(\mathcal{\tilde F}_{\sigma}\mathbb{AF}_{\tau})_zQSHI$, $CDW~(\mathcal{AF}_{\sigma}\mathbb{F}_{\tau})CI$ and $CDWNI$ phase at $U=13$ and $\lambda_{so}=0.6$. The interlayer hopping is $t_l=3,~5,~6~\mbox{and}~8$ respectively.
$E_F$ is the Fermi energy for quarter filling. 

\end{document}